\documentclass[11pt]{article}

\usepackage{amsmath}
\usepackage{amsfonts}
\usepackage{amssymb}
\usepackage{amsthm}
\usepackage{newlfont}
\usepackage{epsfig}
\usepackage{amscd}
\newcommand{\bC}{{\boldsymbol C}}
\newcommand{\bD}{{\boldsymbol D}}
\newcommand{\bE}{{\boldsymbol E}}
\newcommand{\bA}{{\boldsymbol A}}

\newcommand{\ZZ}{{\mathbb Z}}
\newcommand{\KK}{{\mathbb K}}
\newcommand{\FF}{{\mathbb F}}
\newcommand{\LL}{{\mathbb L}}

\newcommand{\be}{\begin{equation}}
\newcommand{\ee}{\end{equation}}
\newcommand{\bes}{\begin{equation*}}
\newcommand{\ees}{\end{equation*}}

\newtheorem{Th}{Theorem}%[section]

\newtheorem{Lem}{Lemma}

\newcommand{\diag}{\mathrm{diag}}

\begin{document}

\title{On pattern structures of the N-soliton solution \\
of the discrete KP equation  over a finite field.}

\author{Mariusz Bia{\l}ecki$^{1,2,}$\footnote{Position $^{1}$ is
thanks to Postdoctoral Fellowship for Foreign Researchers of Japan
Society for the Promotion of Science; $^{2}$ is permanent address.
{\tt email: bialecki@igf.edu.pl}} \ and Jonathan J C
Nimmo$^{1,}$\footnote{Permanent address: Department of Mathematics,
University of Glasgow,
Glasgow G12 8QQ, UK. {\tt email: j.nimmo@maths.gla.ac.uk}} \\ \\
$^{1}$ Graduate School of Mathematical Sciences, University of Tokyo, \\
3-8-1 Komaba, Tokyo 153-8914, Japan\\
 \\
$^{2}$ Institute of Geophysics, Polish Academy of Sciences\\
ul. Ksi\c{e}cia Janusza 64, 01-452 Warszawa, Poland}

\maketitle

\begin{abstract}
The existence and properties of coherent pattern in the multisoliton
solutions of the dKP equation over a finite field is investigated.
To that end, starting with an algebro-geometric construction over a
finite field, we derive a ``travelling wave" formula for $N$-soliton
solutions in a finite field. However, despite it having a form
similar to its analogue in the complex field case, the finite field
solutions produce patterns essentially different from those of
classical interacting solitons.
\end{abstract}

%\subjclass[2000]{14H70, 37K10, 37B15}
PACS 2003: 02.30.Ik, 02.10.De, 05.45.Yv

Keywords: {integrable systems over finite fields;
N-soliton solution; integrable cellular automata; discrete KP equation}.

\section{Introduction}

There are many diverse methods in the modern theory of integrable
systems. One of them, an algebro-geometric approach
\cite{Krich-alg-geo,Krich-discr}, was applied to obtain solutions of
discrete soliton equations over finite fields \cite{DBK, BD-KP,
Bia-1dT}. Within this approach efficient tools for finding
algebro-geometric solutions based on hyperelliptic curves of
arbitrary genus were proposed \cite{BD-hyp, Bia-hyp}. These results
are in direct analogy to the complex field case, but there are some
peculiarities. For instance, since finite fields are never
algebraically closed there are many more possibilities for the
construction of breather type solutions (for short discussion see
\cite{DBK}). Also other properties, such as finiteness or cyclic
structure, reflect in the character of solutions.

Our aim here is to discuss the appearance of stable travelling
patterns for a general $N$-soliton solution over a finite field. To
do that we write a general determinant formula \eqref{eq:det} for
$N$-soliton solution in a travelling wave form \eqref{ex:Ntrav}. The
formula obtained is analogous to the typical form of a soliton
solution of a Hirota bilinear equation (see e.g. \cite{MJD-book},
page 23). Since the vacuum solution in this setting is $\tau \equiv
1$, it is necessary to make a slight modification of the
algebro-geometric construction presented in \cite{BD-KP}.

The determinant form of the solutions of the dKP equation was
already investigated \cite{Ohta}. Since the determinant formula is
neither a Casorati nor a discrete Gram type determinant, we
provide a direct link to the travelling wave form without referring
to the previous work. Moreover, even though we are mainly concerned
with finite fields, we point out the calculations performed are
valid for arbitrary fields.

There is one more reason to transform the finite field solutions
into travelling wave form. There exists a well established systematic
procedure for deriving soliton cellular automata starting from
discrete soliton equations in Hirota form \cite{TTMS, MSTTT}. Since
solutions over a finite field could be interpreted also as cellular
automata, we need a convenient form to investigate relationships
between them.

This article is organized as follows. In Section \ref{sec:AGconst}
we recall and transpose some results from an algebro-geometric
construction of finite field valued solutions of the discrete KP
equation with explicit determinant formula for $N$-soliton solutions
\cite{BD-KP} to the case of dKP with arbitrary coefficients
\eqref{eq:dKPc}. In the next section we prove that the $N$-soliton
solution can be rewritten in travelling wave form. As a remark we
also give an alternative proof which explains why only pairwise
interaction terms appear in the $N$-soliton solutions which appear
as substitution operations for matrix elements. In Section
\ref{sec:EX} we discuss the patterns produced by finite fields
solitons and give some examples. The main conclusion is that
travelling wave patterns in $N$-soliton solutions obtained by the
algebro-geometric approach are generally absent for $N>2$.

\section{Algebro-geometric construction of solutions of dKP equation over finite fields
with simple vacuum.}  \label{sec:AGconst}

A finite field version of an algebro-geometric construction of
solutions $\tau: \ZZ^{3} \rightarrow \FF$  for the discrete KP
equation \be \label{eq:KP} (T_1 \tau) (T_{23} \tau) -  (T_2 \tau)
(T_{13} \tau) + (T_3 \tau)( T_{12} \tau) = 0 \ee was studied in
\cite{BD-KP}. By $T_i$ we denoted here a shift operator in a
variable $n_i$, for example $T_2
\tau(n_1,n_2,n_3)=\tau(n_1,n_2+1,n_3)$. If we prefer to have vacuum
solution $\bar \tau (n_1,n_2,n_3) \equiv 1 $ then we need  to
consider the dKP equation with coefficients $Z_i \in \FF$, i.e. \be
\label{eq:dKPc}
 Z_1 (T_1 \bar\tau) (T_{23} \bar\tau) - Z_2 (T_2 \bar\tau) (T_{13} \bar\tau) +
Z_3 (T_3 \bar\tau)( T_{12} \bar\tau) = 0 \ee with the constraint \be
\label{coefconstr} Z_1 - Z_2 +Z_3 = 0. \ee Note that a
correspondence between $\tau$ and $\bar \tau$ is achieved by taking
\be \label{eq:tautau} \bar\tau = (Z_3/\delta)^{-n_1 n_2}
(Z_1/\delta)^{-n_2 n_3}(Z_2/\delta)^{-n_1 n_3}\tau \ee for nonzero
constants $Z_1, Z_2, Z_3$ and any nonzero $\delta \in \FF$.

It follows from \cite{BD-KP}, in the case of purely soliton
solutions (i.e.\ from a curve of genus $g=0$), that $N$-soliton
solutions can be expressed in a determinant form (Theorem
\ref{th:det} below). As a component of this formula we need a vacuum
solution of \eqref{eq:KP} \be \tau_0 = ({A_1-A_2})^{n_1 n_2}
({A_1-A_3})^{n_1 n_3}({A_2-A_3})^{n_2 n_3}, \ee where $A_i \in \FF
$, and auxiliary functions $\phi_{\alpha}^{0}$, $\alpha=1,2,...N$,
in the form \be \label{pomfi} \phi_{\alpha}(t)=
\frac{1}{t-C_{\alpha}} \prod_{k=1}^{3} \left(
\frac{t-A_k}{C_{\alpha} - A_k} \right)^{n_k}. \ee Parameters
$C_\alpha$, where $\alpha=1,\dots,N$, may take value in some finite
algebraic extension $\LL \supset  \KK$, but they are constrained to
the $\KK$-rationality conditions $\forall \sigma\in  G({\LL}/\KK),
\quad \sigma(C_\alpha) = C_{\alpha^\prime}$. By $G(\LL/\KK)$ we
denoted the Galois group, i.e. the group of automorphisms of $\LL$,
with fixed field $\KK$. Similarly, $N$ pairs $D_\beta, E_\beta\in
\LL$, for $\beta=1,\dots,N$, satisfy the $\KK$-rationality
conditions $\forall \sigma\in  G({\LL}/\KK): \quad
\sigma(\{D_\beta,E_\beta\})=\{D_{\beta^\prime},E_{\beta^\prime}\}.$
These conditions give rise to some generalisation of breather type
solutions (see \cite{DBK}). We assume all parameters in the
construction are distinct. Finally, denote by $\phi_{\bA} (\bD,\bE)$
the $N \times N$ matrix with element in row $\beta$ and column
$\alpha$ given by \be \label{phient} [ \phi_{\bA} (\bD,\bE
)]_{\alpha \beta} = \phi_{\alpha} (D_{\beta}) - \phi_{\alpha}
(E_{\beta}). \ee

\begin{Th} \label{th:det}
The function $\tau(n_1,n_2,n_3)$ given by \be \label{eq:det} \tau =
\tau_0 \cdot \gamma \det \phi_{\bA} (\bD,\bE), \ee where $\gamma \in
\FF$ is some constant, is the $\FF$-valued $N$-soliton solution of
the discrete KP equation \eqref{eq:KP}.
\end{Th}

In the case of dKP equation \eqref{eq:dKPc}, it follows from
\eqref{eq:tautau} that a vacuum function $\bar\tau_0$  can be chosen
in the form \bes \bar\tau_0 = \left(\frac{A_1-A_2}{Z_3}\right)^{n_1
n_2} \left(\frac{A_1-A_3}{Z_2}\right)^{n_1
n_3}\left(\frac{A_2-A_3}{Z_1}\right)^{n_2 n_3}. \ees Then for
parameters $ {A_1-A_2}={Z_3}, \quad {A_1-A_3}={Z_2}, \quad
{A_2-A_3}={Z_1} $ we have $\bar\tau_0 \equiv 1$ and \be
\label{eq:sol} \bar \tau =  \det \phi_{\bA} (\bD,\bE), \ee is the
$\FF$-valued $N$-soliton solution of the equation \eqref{eq:dKPc}.

\noindent {\bf Remark.} The same result can also be derived using
the general algebro-geometric construction of solutions of the
version of the dKP equation given in \eqref{eq:dKPc}. To do this,
the wave function $\psi$ is given by the Definition 1 in
\cite{BD-KP}, but with a different definition of the expansions of
$\psi(n_1,n_2,n_3)$ at $A_i$, namely, for $i,j,k=1,2,3$ with $i\ne
j\ne k\ne i$,
$$
\psi (n_1,n_2,n_3) = t_i^{n_i} \sum_{s=0}^{\infty} Z_j^{n_k}
Z_k^{n_j}\bar\zeta^{(i)}_s(n_1,n_2,n_3)  t_i^s,
$$
where $t_i$ are fixed $\KK$-rational local parameters at $A_i$. The
linear equation for $\psi$ in the general case is of the form
\begin{equation*}
(T_i \psi - T_j \psi)+ Z_k \frac{T_j\bar\zeta^{(i)}_
0}{\bar\zeta^{(i)}_0} \psi = 0. \label{eq:psic}
\end{equation*}
The remaining part of the construction is the same as \cite{BD-KP}.

\section{Travelling waves form for the $N$-soliton solution} \label{sec:TWform}

In this section we wish to transform the soliton solutions given in
\eqref{eq:det} into an equivalent but more convenient form. In doing
this we will use the fact, referred as a gauge invariance, that for
any solution $\tau$ of the dKP equation, $\tau'=\alpha^{n_1}
\beta^{n_2} \gamma^{n_3} \delta  \cdot \tau $ is also a solution for
any constant $\alpha, \beta, \gamma, \delta$. We write
$\tau'\simeq\tau$ if $\tau'$ can be obtained from $\tau$ using this
gauge invariance.

\begin{Lem}\label{lem: inv prod}
Let $M(\mathbf x,\mathbf y)$, where $\mathbf x=(x_{1},\dots,x_{n})$
and  $\mathbf y=(y_{1},\dots,y_{n})$ denote the Cauchy matrix, the
matrix with $(i,j)$th entry $1/(x_{i}-y_{j})$. It is well known that
\begin{equation}\label{eq: cauchy det}
\det M(\mathbf x,\mathbf
y)=\frac{\prod_{p<q}(x_{p}-x_{q})(y_{q}-y_{p})}{\prod_{p,q}(x_{p}-y_{q})}
=\prod_{p<q}\frac{(x_{p}-x_{q})(y_{q}-y_{p})}{(x_{p}-y_{q})(x_{q}-y_{p})}\prod_{p}\frac{1}{(x_{p}-y_{p})}
\end{equation}
Also we have,
\begin{equation}\label{eq: inv prod}
M(\mathbf x,\mathbf y)^{-1}M(\mathbf x,\mathbf z)=
\diag\left(\frac{\prod_{p}(x_{p}-y_{i})}{\prod_{p\ne
i}(y_{p}-y_{i})}\right) M(\mathbf y,\mathbf z)\
\diag\left(\frac{\prod_{p}(y_{p}-z_{i})}{\prod_{p}(x_{p}-z_{i})}\right).
\end{equation}
\end{Lem}

\begin{proof}
Let $\mathbf x_{\hat k}=(x_{1},\dots,x_{k-1},x_{k+1}\dots,x_{n})$.
Then the $(i,j)$th entry in the inverse of $M(\mathbf x,\mathbf y)$
is
\[
[M(\mathbf x,\mathbf y)^{-1}]_{i,j}=(-1)^{i+j}\frac{\det M(\mathbf
x_{\hat\jmath},\mathbf y_{\hat\imath})}{\det M(\mathbf x,\mathbf
y)}=(x_{j}-y_{i})\prod_{p\ne
j}\frac{(x_{p}-y_{i})}{(x_{j}-x_{p})}\prod_{p\ne
i}\frac{(x_{j}-y_{p})} {(y_{p}-y_{i})}.
\]

Further, the $(i,j)$th entry in the product $M(\mathbf x,\mathbf
y)^{-1}M(\mathbf x,\mathbf z)$ is
\begin{align*}
&\frac{\prod_{p}(x_{p}-y_{i})} {\prod_{p\ne i}(y_{p}-y_{i})}
\sum_{k=1}^{n}\frac{\prod_{p\ne i}(x_{k}-y_{p})}
{(x_{k}-z_{j})\prod_{p\ne k}(x_{k}-x_{p})}\\&\qquad=
\frac{\prod_{p}(x_{p}-y_{i})} {\prod_{p\ne i}(y_{p}-y_{i})V(\mathbf
x)\prod_{p}(x_{p}-z_{j})} \sum_{k=1}^{n}(-1)^{n-k}V(\mathbf x_{\hat
k})\prod_{p\ne i}(x_{k}-y_{p})\prod_{p\ne k}(x_{p}-z_{j}),
\end{align*}
where $V(\mathbf x)=\prod_{p>q}(x_{p}-x_{q})$ is the Vandermonde
determinant in variables $x_{1},\dots, x_{n}$ and  $V(\mathbf
x_{\hat k})$ is the Vandermonde determinant in variables
$x_{1},\dots, x_{k-1},x_{k+1},\dots, x_{n}$. In the $(i,j)$th entry,
the sum is of degree $(n-1)(n-2)/2+2(n-1)=n(n-1)/2+n-1$ and we will
factorize it by identifying all of its zeros. First, there are
$n(n-1)/2$ coming when $x_{r}=x_{s}$ for any $r<s$ and $n-1$ from
$y_{r}=z_{j}$ for $r\ne i$. This argument uses the fact that for any
polynomial function $f$ of degree less than $n-1$, one has
$\sum_{k=1}^{n}(-1)^{n-k}V(\mathbf x_{\hat k})f(x_{k})=0$ since the
left hand side is the expansion by the final column of the vanishing
determinant
\[
\begin{vmatrix}
1&x_{1}&\cdots&x_{1}^{n-2}&f(x_{1})\\
\vdots&\vdots&&\vdots&\vdots\\
1&x_{n}&\cdots&x_{n}^{n-2}&f(x_{n})
\end{vmatrix}.
\]
Hence the $(i,j)$th entry is, up to a constant factor,
\[
\frac{\prod_{p}(x_{p}-y_{i})} {\prod_{p\ne i}(y_{p}-y_{i})}
\frac{\prod_{p\ne i}(y_{p}-z_{j})}
{\prod_{p}(x_{p}-z_{j})}=\frac{\prod_{p}(x_{p}-y_{i})} {\prod_{p\ne
i}(y_{p}-y_{i})} \frac1{y_{i}-z_{j}} \frac{\prod_{p}(y_{p}-z_{j})}
{\prod_{p}(x_{p}-z_{j})}.
\]
To verify that it is precisely correct, and hence complete the proof
of \eqref{eq: inv prod}, we observe that, as it should, the above
equals $\delta_{ij}$, the Kronecker delta, when $\mathbf z=\mathbf
y$.
\end{proof}

\begin{Th} \label{th:main}
Let $q$ denote any fixed generator of $\FF^*$ i.e. a multiplicative
subgroup of the finite field $\FF$. The $N$-soliton solution
\eqref{eq:sol} of the dKP equation over a finite field $\FF$ admits
the following form \be \label{ex:Ntrav} \tau' = \sum_{J \subset
\{1,\dots,N\}} (-1)^{\#J}\left(\prod_{i,i' \in J; \ i < i'}
a_{ii'}\right) q^{(\sum_{j\in J} \hat \eta_j )}, \ee where the sum
is taken over all subsets of $\{1,\dots,N\}$  and $\#J$ denotes the
cardinality of $J$. In \eqref{ex:Ntrav}, \be a_{ij}  :=
\frac{(D_i-D_j)(E_i-E_j)}{(D_i-E_j)(D_j-E_{i})} \label{aij}, \ee the
exponents are $\hat \eta_j=\eta_j + \eta^{0}_{j}$ where \be
\label{trexp}
 \eta_j  := \sum_{k=1}^{3} p_j^k n_k ,
 \ee
and the parameters $p_j^k$ and phase constants $\eta^{0}_{j}$ are
defined by \be q^{p_i^k} := \frac{E_i -A_k}{D_i-A_k} \quad
\text{and} \quad q^{\eta^{0}_{i}} :=
\prod_{p=1}^{N}\frac{(C_p-D_i)}{(C_p-E_i)}\prod_{p=1;\,p\ne
i}^{N}\frac{(D_p-E_i)}{(D_p-D_i)}. \ee
\end{Th}

The name travelling wave form comes from the form of the $q^{\hat
\eta_j}$, which are in direct analogy with the usual term $\exp(\vec
k \vec x -\omega t)$ of linear plane waves. Note, that for any
$\tau$ satisfying \eqref{eq:dKPc}, $\tau(1,0,0)$, $\tau(0,1,0)$ and
$\tau(0,0,1)$ depend only on cross ratio of appropriate points on
the projective line. We also point out that to find $p_i^k$ and
$\eta^{0}_{ij}$ we need to solve a discrete logarithm problem.

\begin{proof}
The determinant in \eqref{eq:sol} is
\[
\bar\tau=\det\left(-M
(\bC,\bD)\;\diag\left(\prod_{k=1}^{3}\left(\frac{D_{i}-A_{k}}{C_{i}-A_{k}}\right)^{n_{k}}\right)+M
(\bC,\bE)\;\diag\left(\prod_{k=1}^{3}\left(\frac{E_{i}-A_{k}}{C_{i}-A_{k}}\right)^{n_{k}}\right)\right),
\]
where $M$ denotes the Cauchy matrix as defined in Lemma~\ref{lem:
inv prod}, and so
\[
\bar\tau\simeq \tau':=\det\left(I-M (\bC,\bD)^{-1}M
(\bC,\bE)\;\diag\left(\prod_{k=1}^{3}\left(\frac{E_{i}-A_{k}}{D_{i}-A_{k}}\right)^{n_{k}}\right)\right).
\]
Using \eqref{eq: inv prod} and the fact that for any matrices $P,Q$,
$\det PQ=\det QP$, we get
\[
\tau'=\det(I-M
(\bD,\bE)\;\diag(D_{i}-E_{i})\;\diag(q^{\hat\eta_{i}})).
\]

Now the determinant $\det(P+Q)$, of the sum of two $N\times N$
matrices, may be expressed as the sum over all subsets $J\subset
\{1,\dots,N\}$ of $\det R_{J}$ where the $i$th column of $R_{J}$
equals the $i$th column of $Q$ if $i\in J$ and  otherwise equals the
$i$th column of $P$. Using this fact,
\[
\tau'=\sum_{J\subset \{1,\dots,N\}}(-1)^{\#J}\det
M_{J}(\bD,\bE)\prod_{i\in J}(D_{i}-E_{i})q^{\hat\eta_{i}},
\]
where $M_{J}$ denotes the Cauchy matrix with row and column indices
restricted to $J$. The formula \eqref{ex:Ntrav} now follows
immediately from \eqref{eq: cauchy det}.
\end{proof}

\noindent\textbf{Remark.} We also present here an alternative proof
of Theorem \ref{th:main}. The argument presented shows why there is
only pairwise interaction of solitons.

Starting from \eqref{eq:sol}, for each $i$ one divides the $i$th
column of \eqref{phient} by $\phi_i(D_i)$ and we see that
$\bar\tau\simeq\tau'=\det \Phi $ where \be \label{eq:entries}
[\Phi]_{ij}= \left( \left( \frac{D_i-C_i}{D_j-C_i}\right)  - \left(
\frac{D_i-C_i}{E_j-C_i}\right) q^{\eta_j}  \right) \prod_{k=1}^{3}
\left( \frac{D_j-A_k}{D_i - A_k}\right)^{n_k}, \ee where $\eta_j$ is
given by \eqref{trexp}, is also solution of \eqref{eq:dKPc}.
Further, if we define \be \label{hatPhi} [\hat \Phi]_{ij}:=
 \left( \frac{D_i-C_i}{D_j-C_i}\right) - \left( \frac{D_i-C_i}{E_j-C_i}\right) q^{\eta_j}
\ee it is easy to see that $\det \Phi \simeq \det \hat \Phi$.

We will next show how $\det \hat \Phi$ can be obtained from a vacuum
solution $\det \hat \Phi_0$ where $ [\hat \Phi_0]_{ij}=
 \left( \frac{D_i-C_i}{D_j-C_i}\right)$ by means of \emph {substitution} operations. The key idea we
use in finding coefficients of different powers of $q$ in the
formula \eqref{ex:Ntrav} is observation that, according to
\eqref{hatPhi}, for each term including $q^{\eta_k}$ we should make
the substitution \be \label{replace} \left( \frac{D_i-C_i}{D_k-C_i}
\right) \quad \longrightarrow \quad - \left( \frac{D_i-C_i}{E_k-C_i}
\right) \ee for each $i$, in the $k$th row of $\hat \Phi_0$. Then
$\det \hat \Phi$ can be expressed in terms of determinants obtained
by making appropriate replacements in $\det \hat\Phi_0$.

First notice that \be \label{eq:det0} \det \hat \Phi_0 =  M(-\bC,
-\bD) \ \prod_i (D_i -C_i) =\prod_{i,j=1; \ i<j}^N
\frac{(C_i-C_j)(D_i-D_j)}{(C_i-D_j)(D_i-C_j)}. \ee

Denote by $\Phi_J$ the matrix with replacements \eqref{replace} in
the rows $k \in J $. We first consider the case $J=\{ k \}$ for
fixed $k$. Since $D_k$ appears only in the $k$th row and $k$th
column, then $\Phi_J$ after multiplying $k$th column by $ - \left(
\frac{E_k-C_k}{D_k-C_k} \right)$ becomes the determinant
\eqref{eq:det0} with $D_k$ replaced by $E_k$. Then
\begin{eqnarray*}
\det \Phi_J &=& - \left( \frac{D_k-C_k}{E_k-C_k} \right) \det \hat \Phi_0 \Big|_{D_k \rightarrow E_k}\\
&=& - \left( \frac{D_k-C_k}{E_k-C_k} \right)
\prod_{i,j=1; \ i<j}^N \frac{(C_i-C_j)(D_i-D_j)}{(C_i-D_j)(D_i-C_j)}\Big|_{D_k \rightarrow E_k}\\
&=& - \left( \frac{D_k-C_k}{E_k-C_k} \right) \left( \prod_{j'=1;\
j'\ne k}^N \left( \frac{D_k-D_{j'}}{D_k-C_{j'}} \right)^{-1} \left(
\frac{E_k-D_{j'}}{E_k-C_{j'}} \right)   \right) \det \hat \Phi_0.
\end{eqnarray*}
In the notation of the theorem, the coefficient of $q^{\eta_j}$ is
$\det \Phi_{\{k\}}= - q^{\eta^{0}_k} \det \hat\Phi_0$.

Now we consider coefficient of $q^{ \sum_{k \in J} \eta_{k}}$. This
is the determinant $\det\Phi_J$ obtained from $\det\hat\Phi_0$ by
making replacements \eqref{replace} in rows $k \in J$ for general $J
\subset I$. The number of elements in $J$ is denoted by $\# J$.
Repeating the procedure described above for each $k \in J$ we arrive
at \be \det \Phi_J= (-1)^{\#J} \left( \prod_{k \in J}
\frac{D_{k}-C_{k}}{E_{k}-C_{k}} \right) \det \hat\Phi_0 \Big|_{D_{k}
\rightarrow E_{k}; \ k \in J} . \ee Then we have
\begin{eqnarray*}
 \det \hat\Phi_0 \Big|_{D_{k} \rightarrow E_{k}; \ k \in J}  &=&
\left( \prod_{ \{k_i, k_j\} \subset J}
 \frac{(D_{k_1}-D_{k_2})(E_{k_1}-E_{k_2})}{(E_{k_1}-D_{k_2})(D_{k_1}-E_{k_2})} \right) \\
&\times & \prod_{k \in J} \left( \prod_{j'=1;\ j'\ne k}^N \left(
\frac{D_k-D_{j'}}{D_k-C_{j'}} \right)^{-1} \left(
\frac{E_k-C_{j'}}{E_k-C_{j'}} \right)   \right) \det \hat\Phi_0.
\end{eqnarray*}
The extra terms for pairs $\{k_i,k_j\} \in J$ are present because,
since $k_j \ne k_i$, we also need to perform the replacements $
D_{k_j} \rightarrow E_{k_j}$ in the factor $ \prod_{j'=1;\ j'\ne
k_i}^N \left( \frac{D_{k_i}-D_{j'}}{D_{k_i}-C_{j'}} \right)^{-1}
\left( \frac{E_{k_i}-D_{j'}}{E_{k_i}-C_{j'}} \right) $. This is done
by multiplying by $
\frac{(D_{k_1}-D_{k_2})(E_{k_1}-E_{k_2})}{(E_{k_1}-D_{k_2})(D_{k_1}-E_{k_2})}$.
We point out here the extra corrections are for \emph{pairs}
$\{k_i,k_j\} \in J$ and it is this that is responsible for the
existence of only \emph{pairwise} interaction terms in the formula
\eqref{ex:Ntrav}. So finally, the coefficient of $q^{\sum_{j\in
J}\eta_j}$ is \be \det\Phi_J  =  (-1)^{\#J} \left(\prod_{i,i' \in J;
\ i < i'} a_{ii'}\right) \left(  \prod_{k \in J} q^{\eta_k^{0}
}\right) \det\hat\Phi_0 . \ee After dividing all terms by
$\det\hat\Phi_0$ and collecting them together we obtain formula
\eqref{ex:Ntrav}.

Notice that our considerations are valid for any field of definition
of the parameters $C_i$ and $D_j$.

\paragraph{Remark.} The results obtained above apply also to the Discrete
Analogue of Generalized Toda Equation \cite{Hirota}
$$ Z_1 f_{m_1+1,m_2,m_3} f_{m_1-1,m_2,m_3}
+ Z_2 f_{m_1,m_2+1,m_3} f_{m_1,m_2-1,m_3} + Z_3 f_{m_1,m_2,m_3+1}
f_{m_1,m_2,m_3-1} = 0.$$ The transition from dKP to DAGTE (where we
assume $(m_1+m_2+m_3) \equiv 0 \mod 2$) is given by
$f_{m_1,m_2,m_3}:=\tau (n_1,n_2,n_3)$ where $n_i=\frac{1}{2} (m_i
-m_j -m_k) \quad \text{for} \quad i \ne j \ne k \ne i $, or
equivalently $ m_i = -(n_j + n_k) \quad \text{for} \quad i \ne j \ne
k \ne i. $ In this situation $f_{0,0,0}=\tau(0,0,0)$. With the
variables ${m_1,m_2,m_3}$ the exponent of a travelling wave given by
\eqref{trexp} have similar form, namely $ \eta_i = \sum_{j=1}^{3}
\hat p_{i}^j m_j$ , where $\hat p_{i}^j = \sum_{k=1}^{3} (2
\delta_{jk} - 1 )p_k$ and $\delta_{jk}$ is the Kronecker delta.

\section{Pattern structures in soliton interactions} \label{sec:EX}

So far we have seen that the finite field $N$-soliton solutions we
obtained have the same structure as the complex field case. In
particular, the travelling wave form \eqref{ex:Ntrav} is completely
analogous to the complex field case. Despite this correspondence,
the typical soliton-like interaction pattern is not present in the
finite field case. Below we argue that one cannot expect that an
arbitrary $N$-soliton solution over a finite field will be a
collection of asymptotically separated waves interacting in the way
characteristic of the complex field case.

\begin{figure}[!p]
\noindent Elements of $\FF_{17}$ are represented on the scale below:
from 0 - dark to 16 - light gray.
\begin{center}
{\leavevmode\epsfysize=0.8cm\epsffile{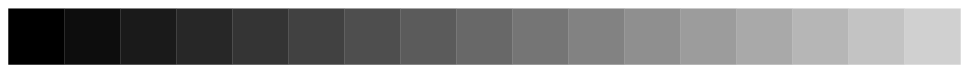} }
\end{center}
\begin{center}
\leavevmode\epsfysize=5cm\epsffile{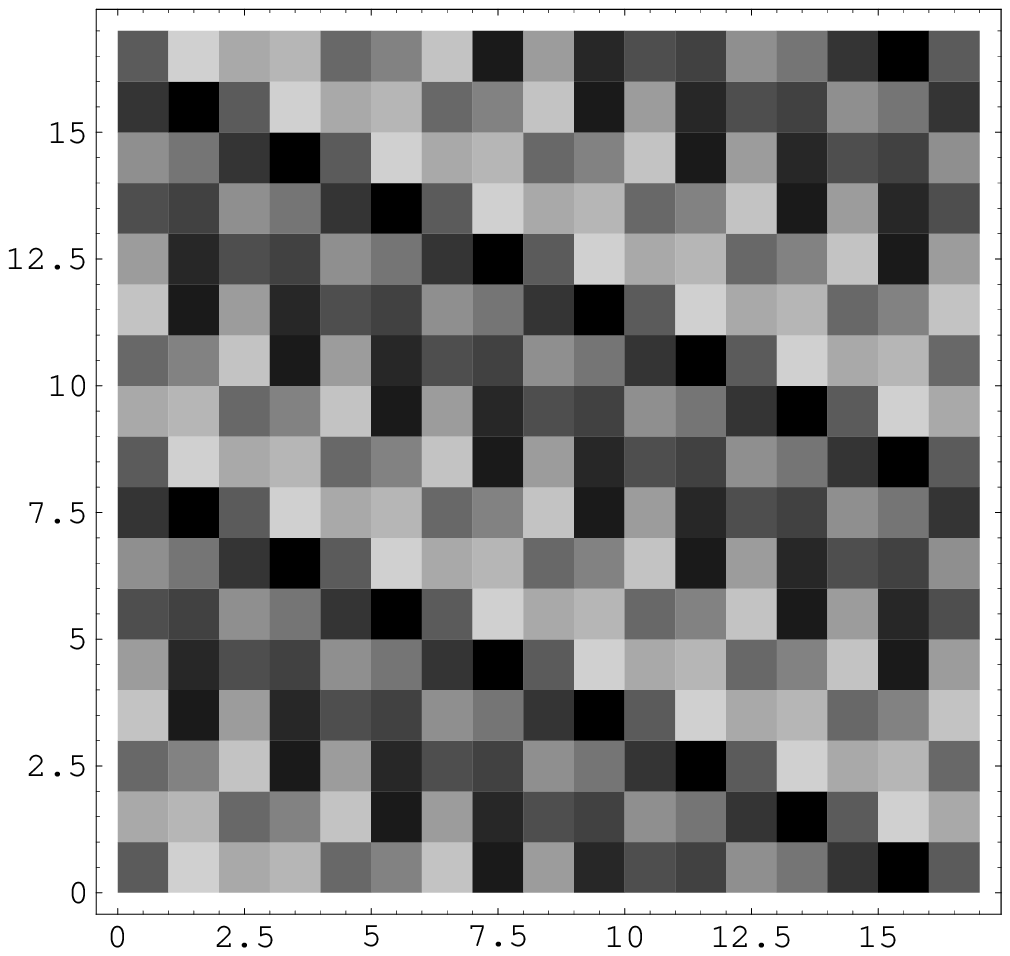} \hspace{0.2cm}
\leavevmode\epsfysize=5cm\epsffile{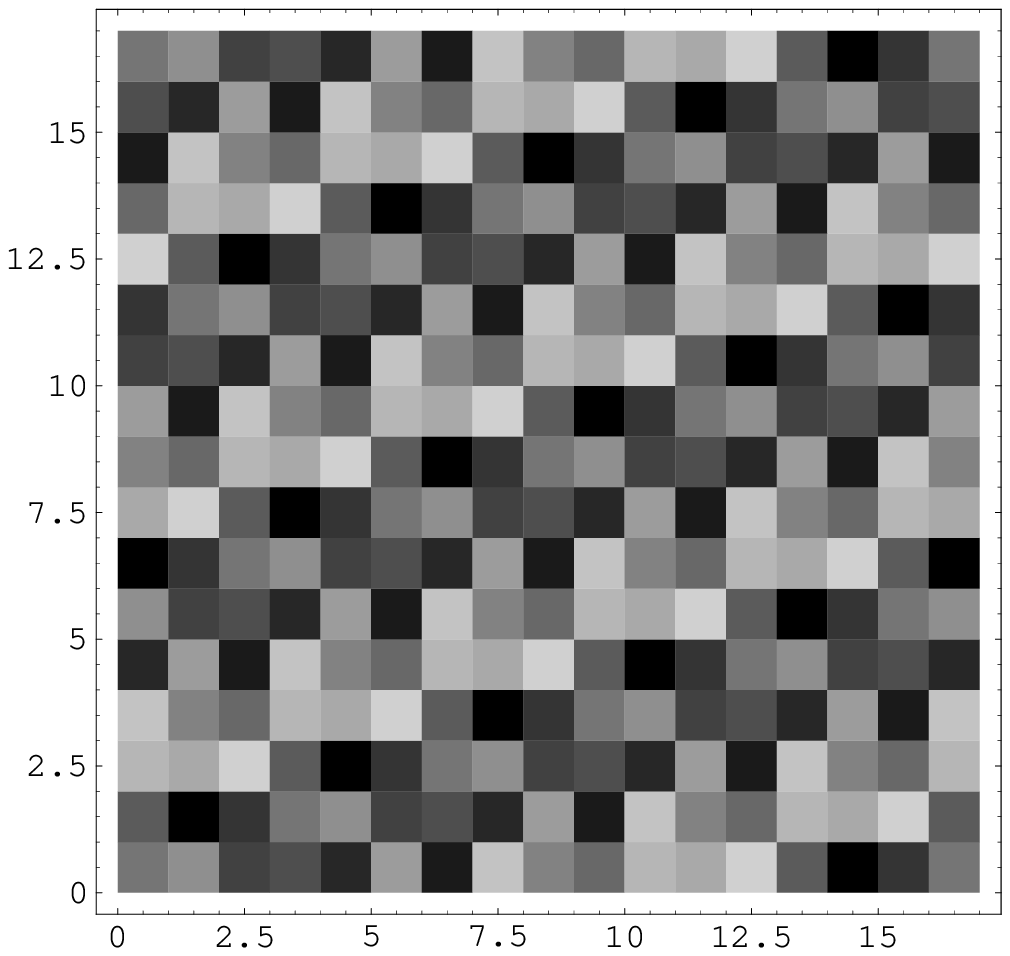} \hspace{0.2cm}
\leavevmode\epsfysize=5cm\epsffile{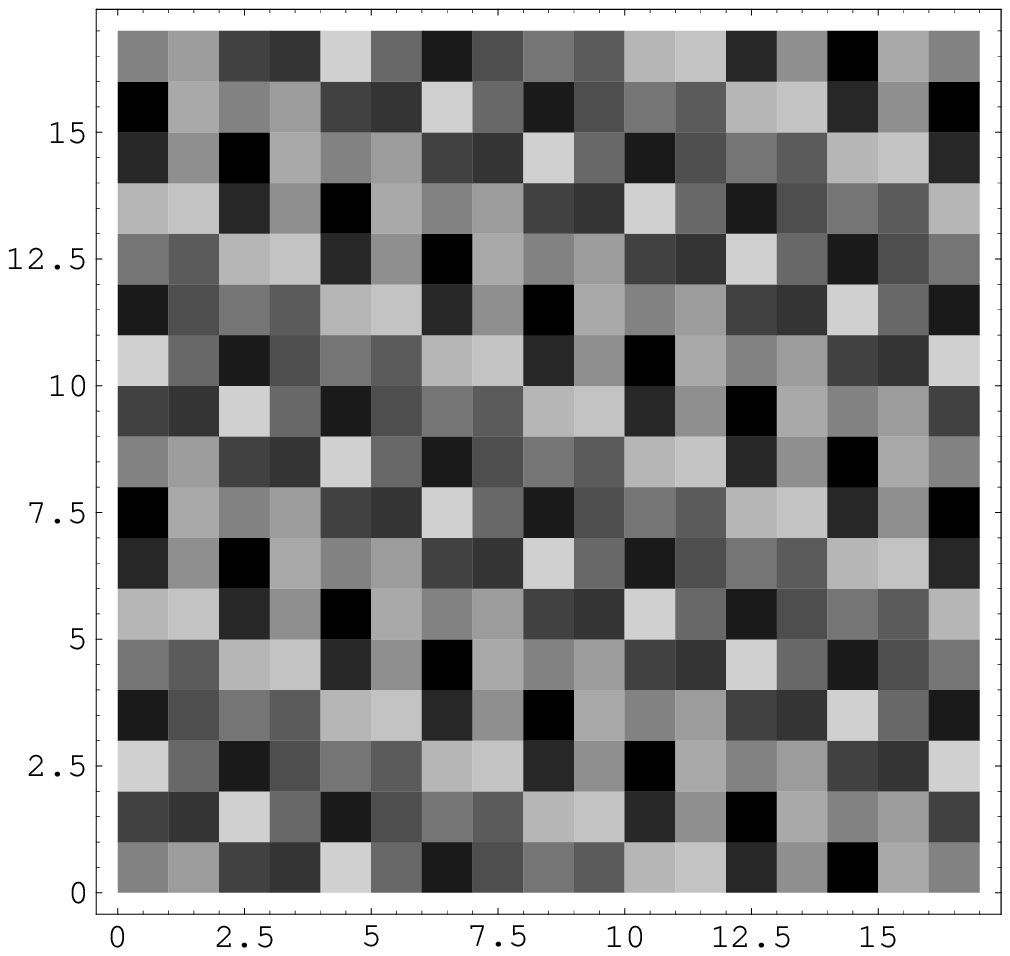}
\end{center}
\caption{ A plot of $\tau(n_1,n_2,n_3)$ function of a three
one-soliton solutions (see {Examples} for details). We fix
$n_1=0$, and  $n_2,n_3 \in \{0,\ldots,16 \}$. The $n_2$ axis is
directed to the right and the $n_3$ axis is directed upwards. }
\label{fig1}
\end{figure}

These differences in the interaction properties follow naturally
from the differences in structure between finite fields and the
complex numbers. First, finite fields have no good (total) ordering
which are consistent for both addition and multiplication and so it
is impossible to find in the finite filed case a direct analogue of
wave amplitude; rather than follow a wave propagation by observing
how its points of maximum amplitude move we may trace only the
propagation of patterns. The appearance of a one-soliton solution
however in the finite field case mimics the usual appearance quite
well, as can be seen on Figure~\ref{fig1}.

The next difference results from the replacement of exponentials in
the complex case by powers of a generator $q$ of the multiplicative
group $\FF^*$ of the finite field $\FF$. Since $q^{|\FF|-1}=1$, this
implies periodicity of $\tau(n_1,n_2,n_3)$ with respect to each
variable $n_i$. As a consequence we can not discuss the asymptotic
behaviour of the solution in usual sense. Moreover, for any given
solution one could restrict analysis to a finite \emph{base cube}
containing all information about the solution, the rest is but
periodic repetition. The length of any edge of a base cube is at
most $|\FF|-1$.

Considering Theorem \ref{th:main}, the $i$th one-soliton component
of the $N$-soliton solution is unchanged by a shift in the lattice
by $\vec n^i=(n^i_1,n^i_2,n^i_3)$ for any  $n^i_1,n^i_2,n^i_3$
satisfying $q^{\eta_i} = 1$, or equivalently, \be \label{periods}
   \eta_i  = \sum_{k=1}^{3} p_i^k n^i_k \equiv 0 \mod (|\FF|-1).
\ee Since expression \eqref{ex:Ntrav} contains $q^{\eta_i}$ for $i
\in  \{1,2,\ldots,N\}$, a period vector $\vec n=(n_1,n_2,n_3)$ for
this solution should be a common solution of \eqref{periods} for all
$i$. In general, it is impossible to find a nonzero solution for $N
\geq 3$ and it means there is no additional structure within the
base cube in this case. An example of such a $3$-soliton solution is
shown in Figure \ref{fig3} and the structure of a base cube is
discussed below in more detail.

\paragraph*{Examples.}

Fix the finite field to be $\FF=\FF_{17}$, i.e.{} the field of integers modulo $17$. As a
generator of $\FF^*$ we choose $q=3$. Let us fix $A_1 =7$, $A_2 = 4$
and $A_3 = 3$ so that the coefficients in the dKP equation are
${A_2-A_3}={Z_1}=1$, ${A_1-A_3}={Z_2}=4$ and ${A_1-A_2}={Z_3}=3$.

\emph{One-soliton solutions.} In Figure \ref{fig1} we present three
one-soliton solutions of the dKP equation over $\FF$. For the
solution A on the left, denoted by subscript $_1$ we have chosen $
C_1 = 11$, $D_1 = 6$, $E_1 = 9$; for the solution B in the middle,
denoted by $_2$, we have $C_2 = 10$, $D_2 =12$, $E_2 =14$ and
finally $C_3 = 8$, $D_3 =13$, $E_3 =15$ for the solution C on the
right. The respective parameters in the exponent \eqref{trexp} are
$$\vec p_1=(p_1^{1},p_1^{2},p_1^{3}) = (6,7,14),
 \quad \vec p_2=(p_2^{1},p_2^{2},p_2^{3}) = (6,9,5),
\quad \vec p_3=(p_3^{1},p_3^{2},p_3^{3}) = (11,5,10).$$ Thus periods
in variables $n_1,n_2,n_3$ are: $8,16,8$ for the soliton $_1$,
$8,16,16$ for the soliton $_2$ and $16,16,8$ for $_3$. We fixed $n_1
=0$, since increasing any variable by 1 results in shift in the
other two variables and so the plots for any value of $n_1$ is
simply a translation of the plots we present. This comes from the
fact that in general there exists a nonzero solution in two
variables $n_j, n_k$ of the single equation \eqref{periods} with the
third variable $n_i$ being fixed ($i,j,k\in\{1,2,3\}$, $i\ne j\ne k
\ne i$). The special case is if $p^i$ is not a linear combination of
$p^j$ and $p^k$ with coefficients from $\ZZ \mod (|\FF|-1)$. (This
could happen for instance if $p^j$ and $p^k$ are zero divisors of
$|\FF|-1$.) In the case presented in Figure \ref{fig1}, period
vectors might be chosen to be $\vec n_1=(1,0,3)$ for A,
 $\vec n_2=(1,0,2)$ for B and $\vec n_3=(1,1,0)$ for C.
Further, the period vectors $\vec n$ for these solutions in the
plane $n_2,n_3$ are  $\vec n_{1a} = \vec n_{3a} = (0,2,-1)$, $\vec
n_{1b} = \vec n_{3b} = (0,0,8)$,  $\vec n_{2a} = (0,3,1)$ and  $\vec
n_{2b} = (0,-1,5)$. (All periods vectors are not uniquely
determined.)

\begin{figure}[!p]
%\noindent Elements of $\FF_{17}$ are represented on the scale below:
%from 0 - dark to 16 - light gray.
%\begin{center}
%{\leavevmode\epsfysize=0.8cm\epsffile{skala.eps} }
%\end{center}
\begin{center}
\leavevmode\epsfysize=5cm\epsffile{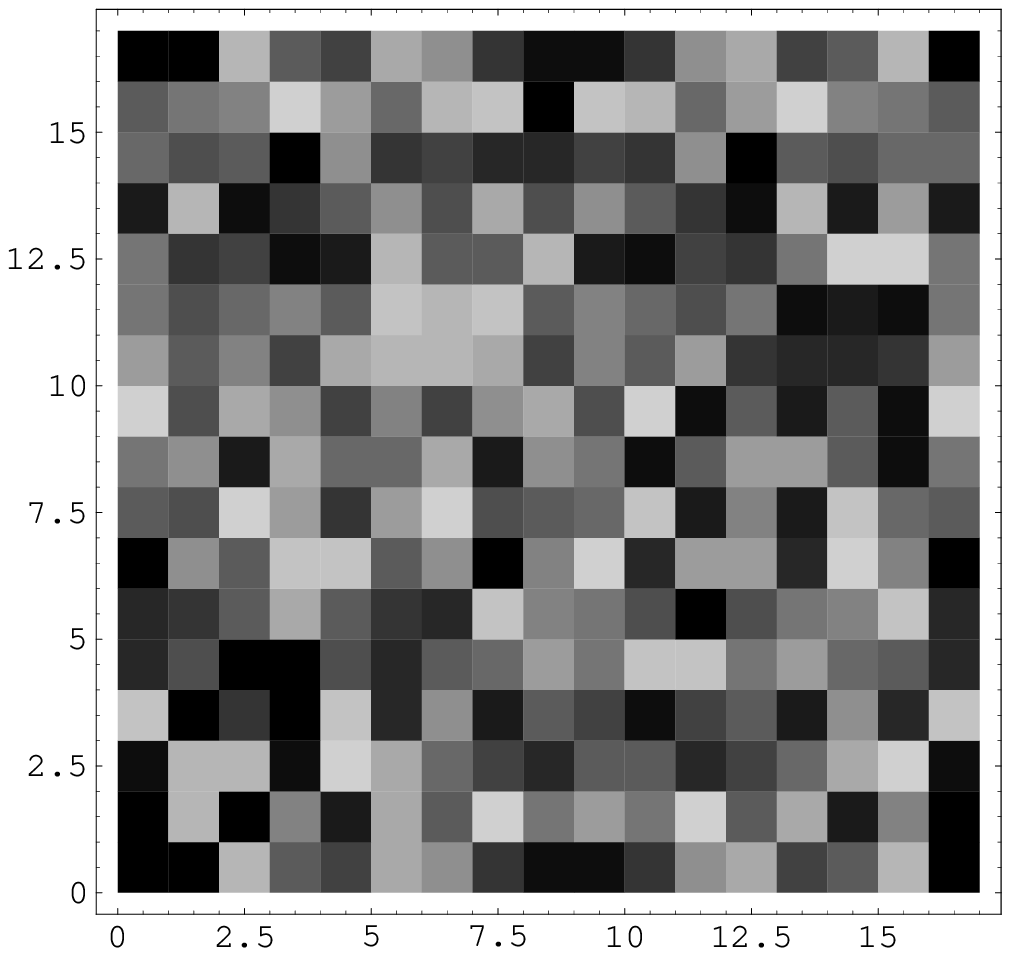} \hspace{0.2cm}
\leavevmode\epsfysize=5cm\epsffile{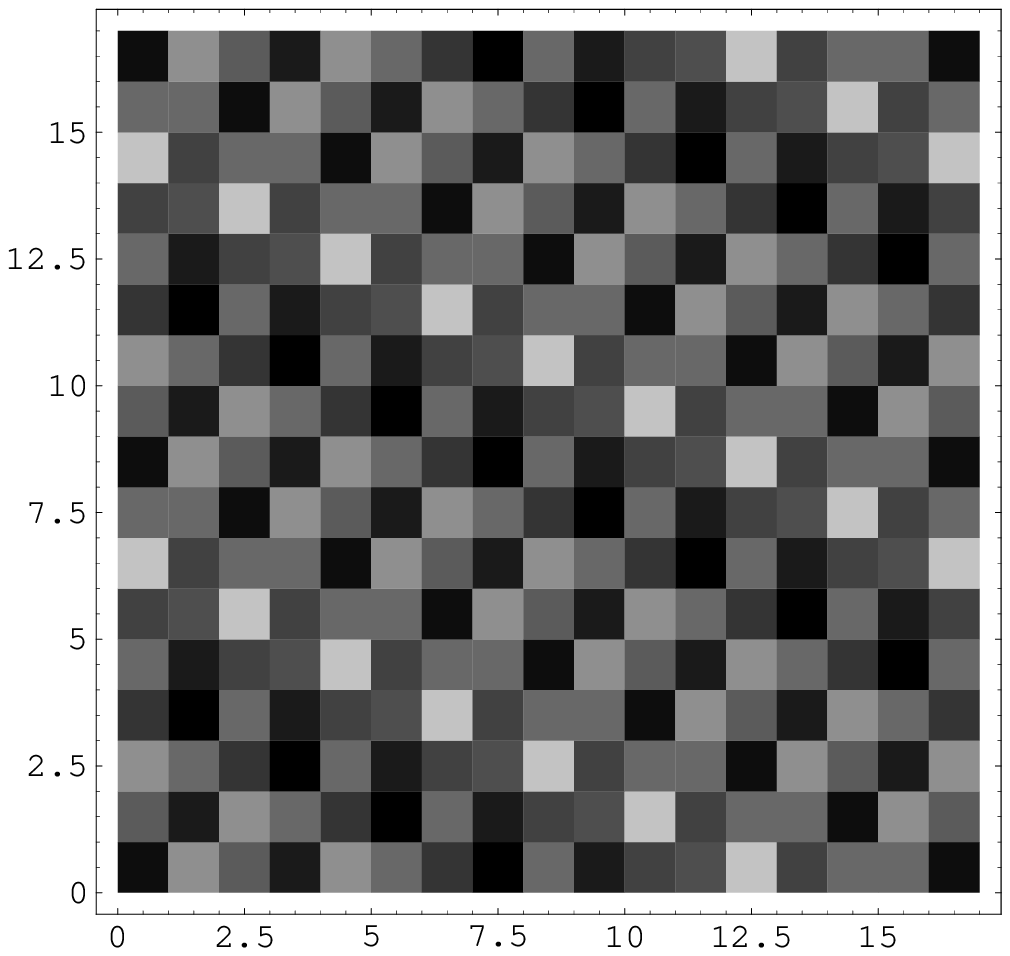} \hspace{0.2cm}
\leavevmode\epsfysize=5cm\epsffile{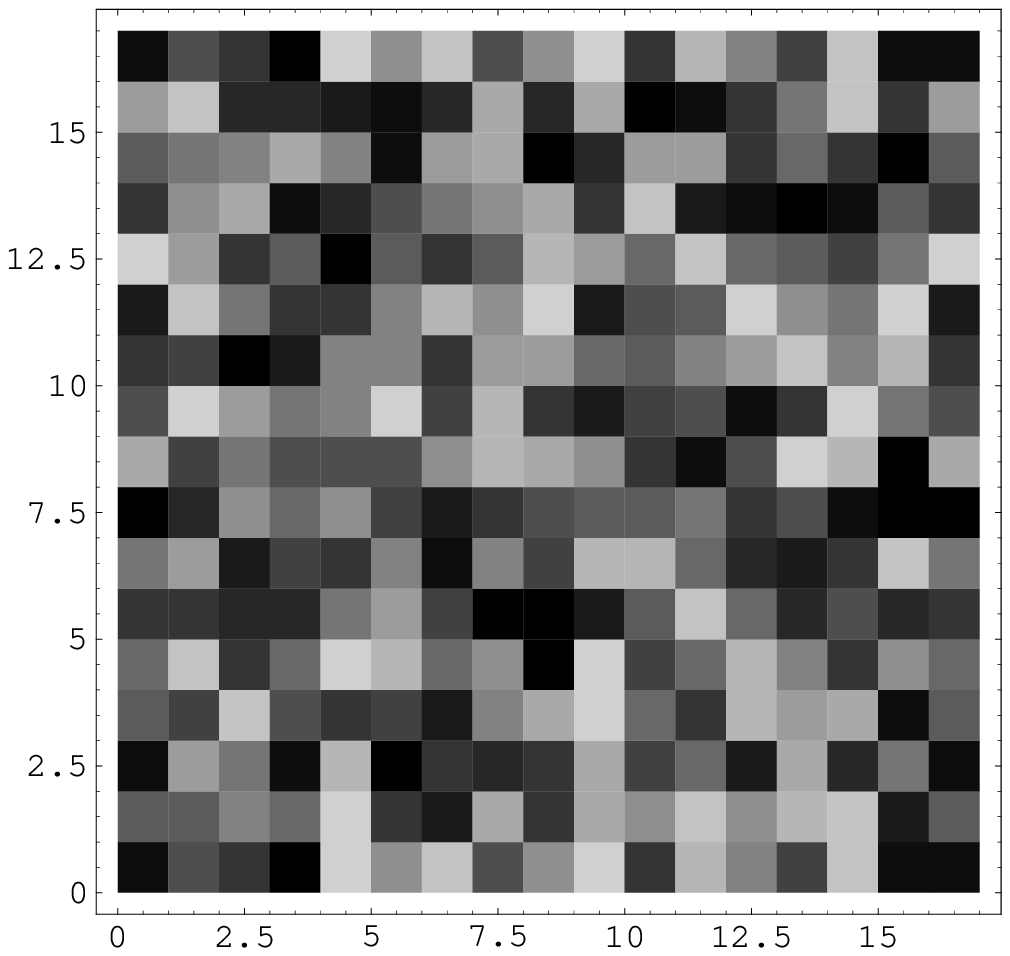}
\leavevmode\epsfysize=5cm\epsffile{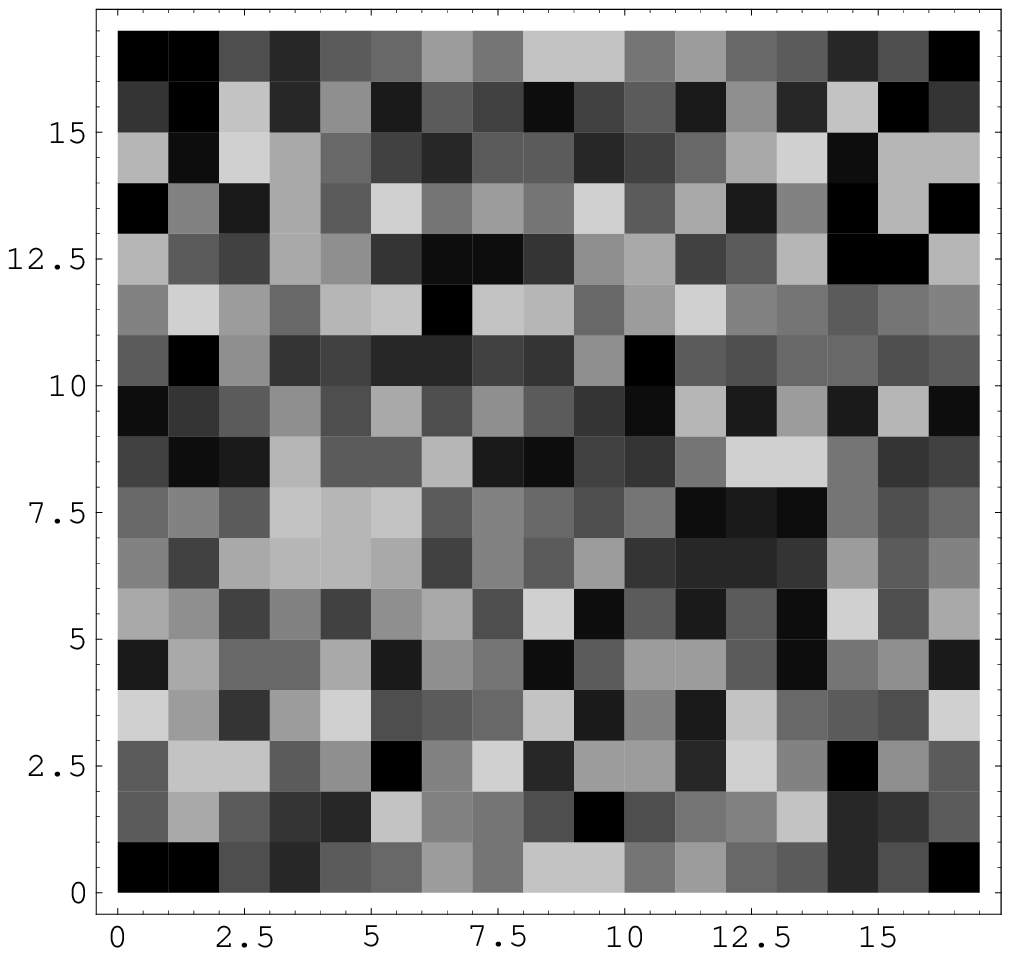} \hspace{0.2cm}
\leavevmode\epsfysize=5cm\epsffile{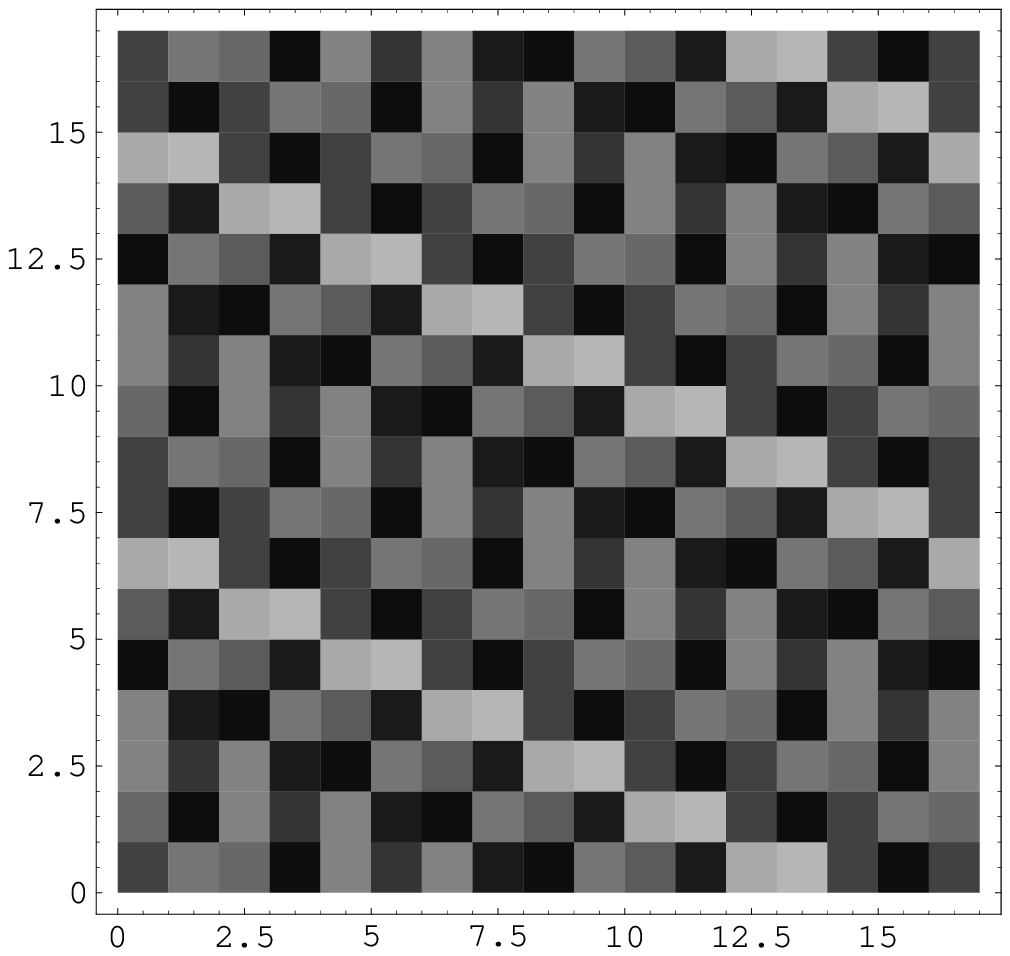} \hspace{0.2cm}
\leavevmode\epsfysize=5cm\epsffile{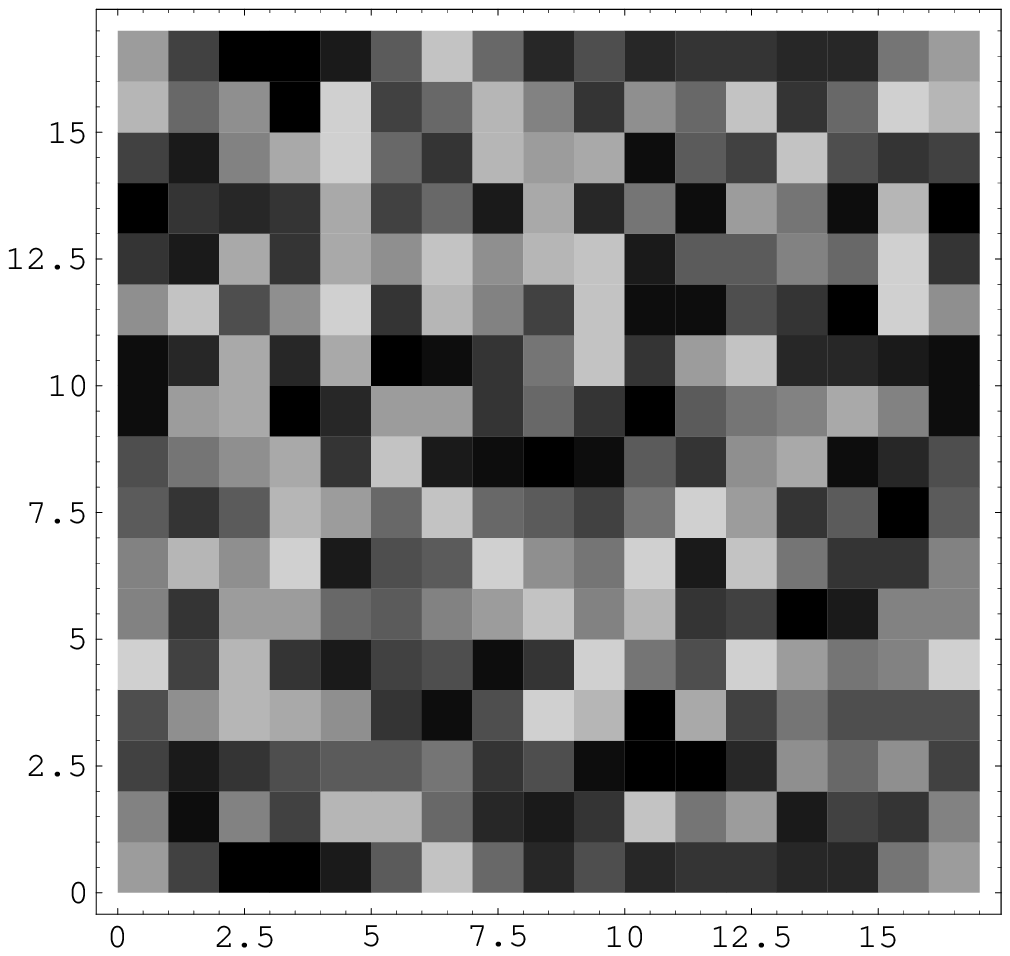}
\end{center}
\caption{ A plot of $\tau(n_1,n_2,n_3)$ for the three two-soliton
interactions (AB, AC, BC) of one-soliton solutions presented in
Figure \ref{fig1}. We fix $n_1=0$, and $n_2,n_3 \in \{0,\ldots,16
\}$. In the second row we have $n_1=1$. The $n_2$ axis is directed
to the right and the $n_3$ axis is directed upwards. } \label{fig2}
\end{figure}

\newcommand{\sspan}{\mathrm{span}}

\emph{Two-soliton solutions.} In Figure~\ref{fig2} the two-soliton
solutions representing the interaction of two of the one-soliton
solutions shown in Figure~\ref{fig1} are presented. The upper line
is for $n_1=0$ while the lower is for $n_1=1$. In the cases A
interacting with B (called AB) and B interacting with C (AC) it is a
simple shift by respective vectors $(1,-1,-4)$ and $(1,-5,-5)$.
(Note: Since $q^{16}=1$, the lines $n=0$ and $n=16$ are identical
but both of them are shown on plots.) For the case AC
there are no such a shift, because the two equations \eqref{periods}
for $i=1$ and $i=3$ for $n_2$ and $n_3$ has no nonzero solution for
any value of $n_1$. This is because $ n_1 \cdot (6,11) \not\in
\ZZ_{16} \cdot (7,5)$. Since $\sspan(\vec n_{1a}, \vec n_{1b}) \cap
\sspan (\vec n_{2a}, \vec n_{2b}) = \{ \vec 0\}$ there are no
nonzero period vectors in the plane $n_2, n_3$ for either AB or BC.
On the contrary, for AC the period vectors in this plane are exactly
the same as for soliton A and C (since they are the same). Similarly
one could examine other planes obtaining plane period vectors
$(0,0,8)$ for AC, $(8,0,0)$ and $(4,8,0)$ for AB and none for BC.
Periods for AB are $(8,16,16)$, for AC are $(16,16,8)$ and for BC
are the maximal $(16,16,16)$. Notice that BC shows that it is
possible for a two-soliton solution to have no structure within the
base cube.

\emph{A three-soliton solution.} In this case there are three
equations \eqref{periods} for three variables, so in general there
are no nonzero solutions. The example presented in Figure \ref{fig3}
using parameters in A, B and C is of this kind. Because of this,
there are no extra period vectors and so this solution has no
structure within a base cube.

\begin{figure}[!t]
\noindent Elements of $\FF_{17}$ are represented on the scale below:
from 0 - dark to 16 - light gray.
\begin{center}
{\leavevmode\epsfysize=0.8cm\epsffile{skala.eps} }
\end{center}
\begin{center}
\leavevmode\epsfysize=5cm\epsffile{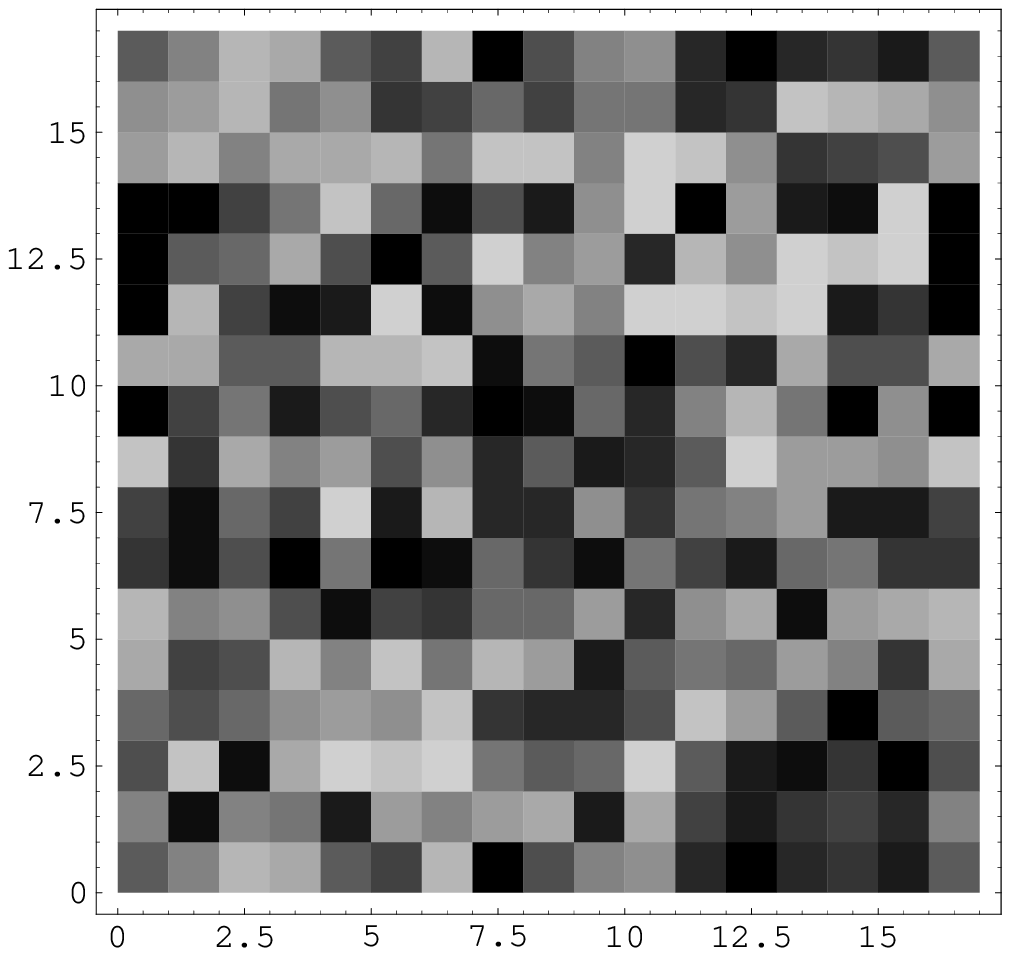} \hspace{0.2cm}
\leavevmode\epsfysize=5cm\epsffile{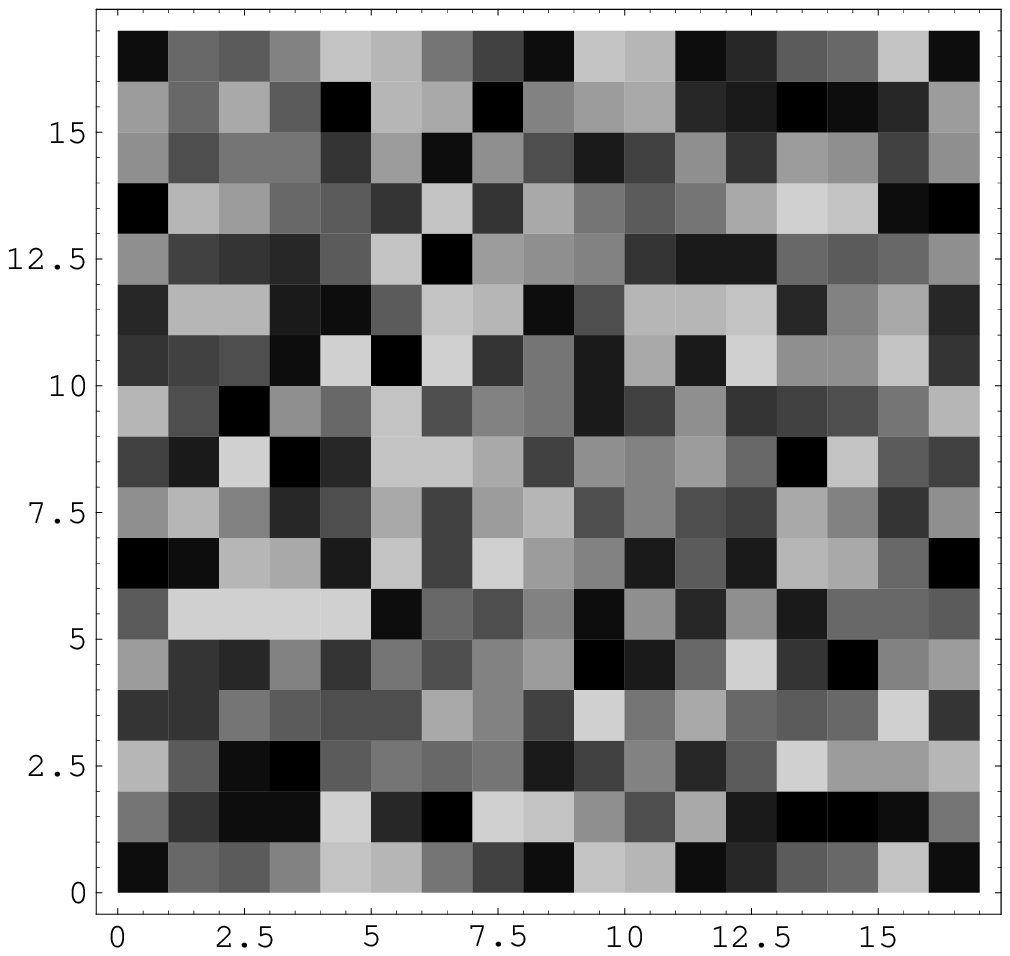} \hspace{0.2cm}
\leavevmode\epsfysize=5cm\epsffile{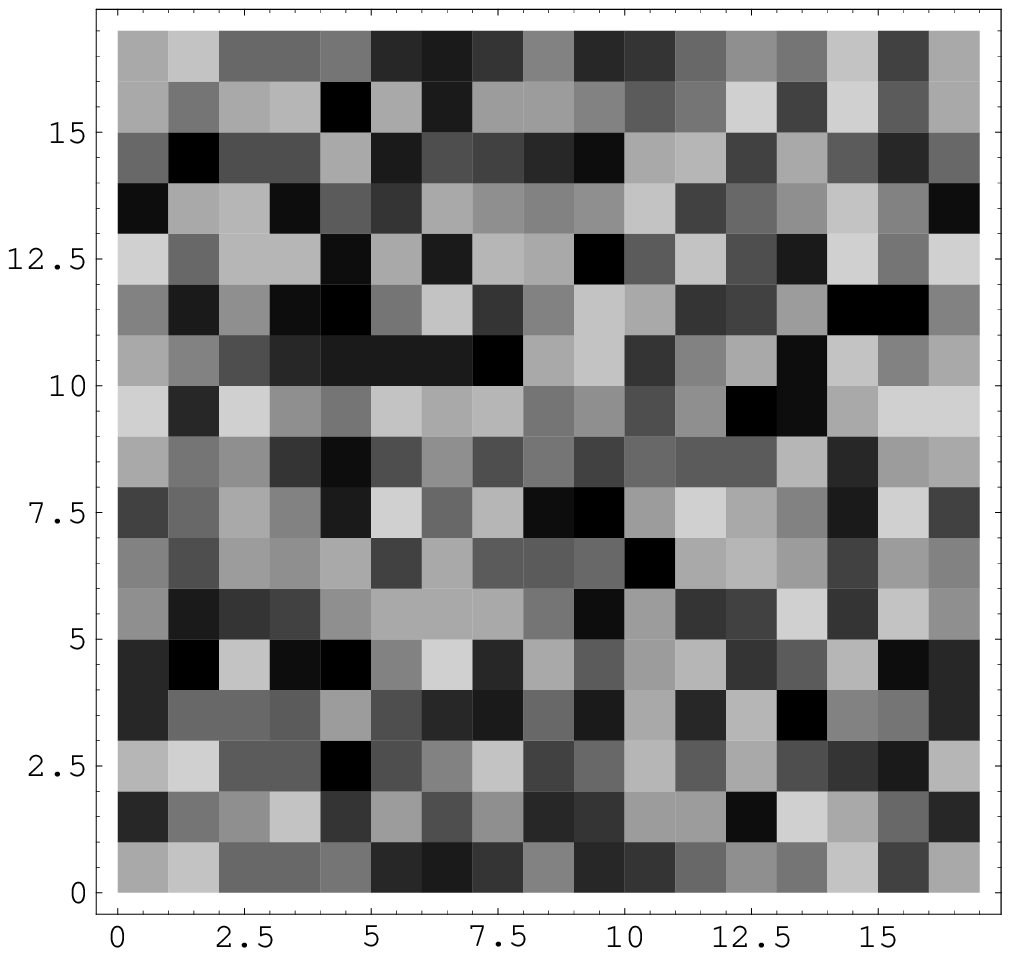}
\leavevmode\epsfysize=5cm\epsffile{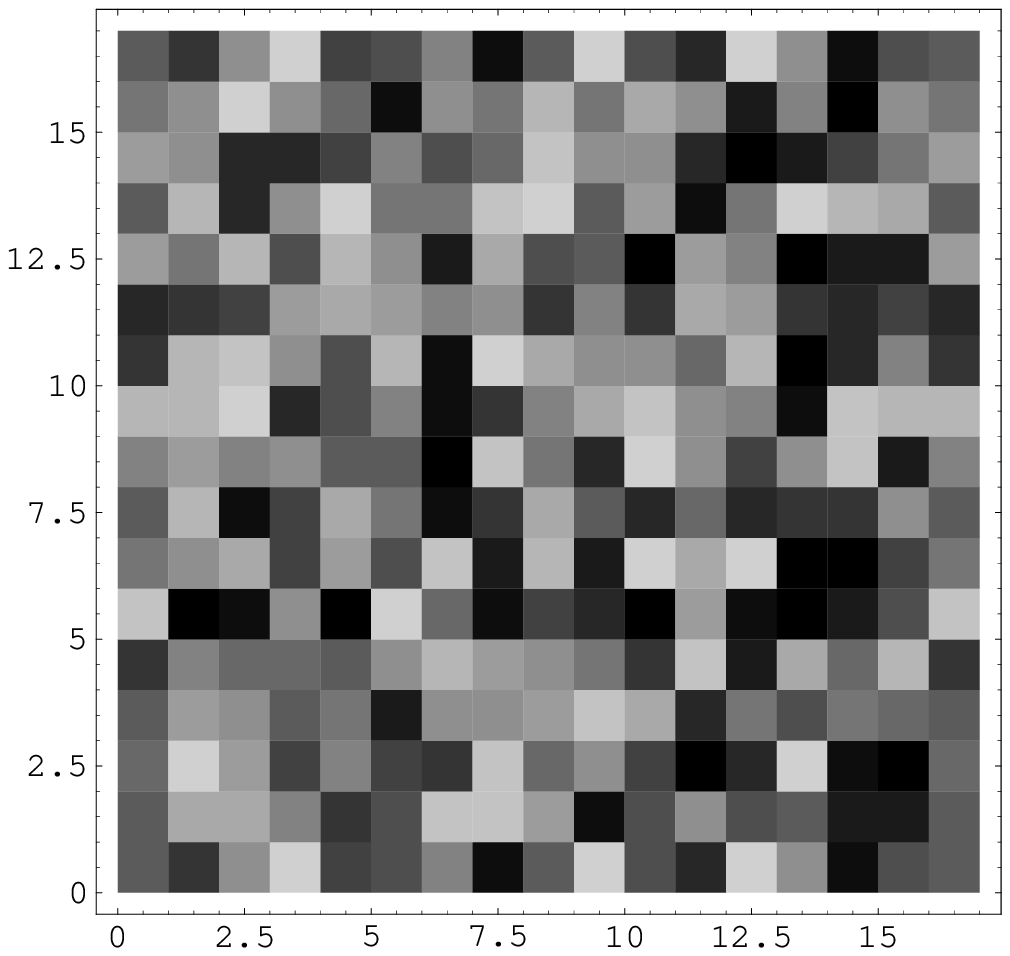} \hspace{0.2cm}
\leavevmode\epsfysize=5cm\epsffile{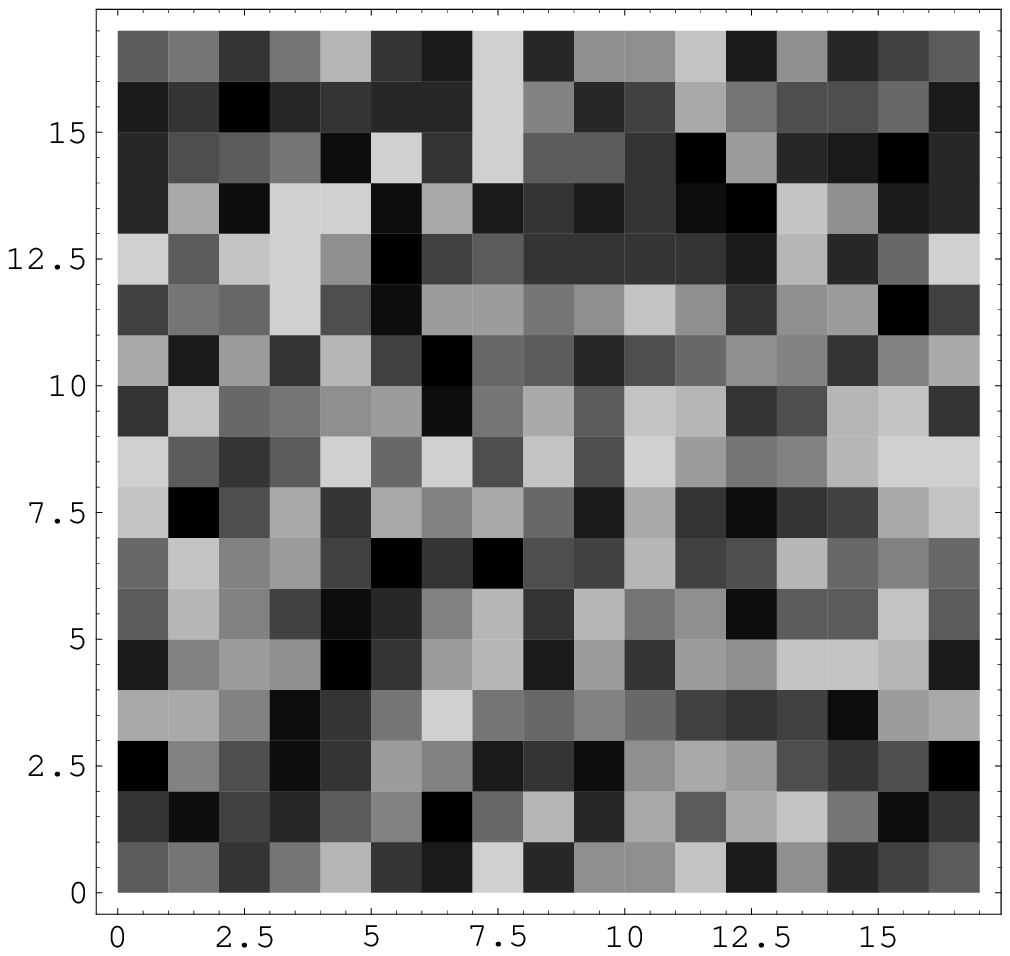}
\end{center}
\caption{ A three-soliton solution $\tau(n_1,n_2,n_3)$ being the
solitonic sum of those from Figure \ref{fig1} for $n_1=0,1,2,4$ and
$8$. Axes: $n_2$ directed to the right, $n_3$ directed upward. }
\label{fig3}
\end{figure}

It is clear that this is also the generic situation for the
interaction of $N$-soliton solutions obtained by the
algebro-geometric approach for $N>2$. In short, we have seen that
the interaction of three or more finite field soliton solutions of
the dKP equation has no more structure in the base cube than random
data. Given a finite field $\FF$ of sufficient size, it would seem
to be computationally impractical to attempt to reconstruct the
parameters $C_\alpha$, $D_\alpha$ and $E_\alpha$ which define the
solution, from the seemingly random data that they generate. This
leads one to imagine possible applications of such solutions in
encrypted data transmission. In conclusion, we note that
applications of elliptic curves over finite fields have already been
studied for some time \cite{Koblitz,Miller}

\section*{Acknowledgements}

The authors thanks Professors T. Tokihiro and R. Willox for creating
a friendly environment for research at the Graduate School of
Mathematical Sciences, University of Tokyo and for their overall
help.

M.B. is partially supported by Grant-in-Aid of the JSPS
Postdoctoral Fellowship for Foreign Researchers (ID No.P 05050) and
by Polish Ministry of Science and Information Society Technologies
(Grant no. 1~P03B~01728).

%\newpage
%\providecommand{\bysame}{\leavevmode\hbox
%to3em{\hrulefill}\thinspace}
%\providecommand{\MR}{\relax\ifhmode\unskip\space\fi MR }
%\providecommand{\MRhref}[2]{%
%  \href{http://www.ams.org/mathscinet-getitem?mr=#1}{#2}
%} \providecommand{\href}[2]{#2}


\begin{thebibliography}{10}



\bibitem{Bia-1dT}
M.~Bia{\l}ecki, \emph{Integrable 1{D} {T}oda cellular automata}, J.
Nonl. Math.
  Phys. \textbf{12}, Suppl. 2, (2005), 28--35.



\bibitem{Bia-hyp}
M.~Bia{\l}ecki, \emph{Integrable {KP} and {K}d{V} cellular automata out of
a
  hypereliptic curve}, Glasgow Math. J. \textbf{47A} (2005), 33--44.



\bibitem{BD-KP}
M.~Bia{\l}ecki and A.~Doliwa, \emph{The discrete {KP} and {KdV}
equations over
  finite fields}, Theor. Math. Phys. \textbf{137(1)} (2003), 1412--1418.



\bibitem{BD-hyp}
M.~Bia{\l}ecki and A.~Doliwa, \emph{Algebro-geometric solution of the discrete {KP}
equation over a
  finite field out of a hyperelliptic curve}, Commun. Math. Phys. \textbf{253}
  (2005), 157--170.



\bibitem{DBK}
A.~Doliwa, M.~Bia{\l}ecki, and P.~Klimczewski, \emph{The {H}irota
equation over
  finite fields: algebro-geometric approach and multisoliton solutions}, J.
  Phys. A: Math. Gen. \textbf{36} (2003), 4827--4839.



\bibitem{Hirota}
R.~Hirota, \emph{Discrete analogue of a generalized {Toda}
equation}, J. Phys.
  Soc. Jpn. \textbf{50} (1981), 3785--3791.


\bibitem{Koblitz} N.~Koblitz, \emph{Elliptic curve cryptosystems},
Math. Comp. \textbf{48} (1987), 203--209.



\bibitem{Krich-alg-geo}
I.~M. Krichever, \emph{Methods of algebraic geometry in the theory
of nonlinear
  equations}, Uspiekhi Mat. Nauk \textbf{32:6} (1977), 183--208.



\bibitem{Krich-discr}
I.~M. Krichever, \emph{Algebraic curves and nonlinear difference equations},
Uspiekhi
  Mat. Nauk \textbf{33:4} (1978), 215--216.



\bibitem{MSTTT}
J.~Matsukidaira, J.~Satsuma, D.~Takahasi, T.~Tokihiro, and M.~Torii,
  \emph{Toda-type cellular automaton and its {N}-soliton solution}, Phys. Lett.
  A \textbf{225} (1997), 287--295.



\bibitem{Miller}
V. S. Miller, in {\it Advances in cryptology---CRYPTO '85 (Santa
Barbara, Calif., 1985)}, Lecture Notes in Comput. Sci., 218,
Springer, Berlin, (1986), 417--426.



\bibitem{MJD-book}
T.~Miwa, M.~Jimbo, and E.~Date, \emph{Solitons: differential
equations,
  symmetries and infinite dimensional algebras}, University Press, Cambridge,
  2000.



\bibitem{Ohta}
Y.~Ohta, R.~Hirota, and S.~Tsujimoto, \emph{Casorati and discrete
{G}ram type
  determinant representations of solutions of the discrete {KP} hierarchy},
  J.Phys.Soc.Jpn \textbf{62:6} (1993), 1872--1886.



\bibitem{TTMS}
T.~Tokihiro, D.~Takahashi, J.~Matsukidaira, and J.~Satsuma,
\emph{From soliton
  equations to integrable cellular automata through a limiting procedure},
  Phys. Rev. Lett. \textbf{76} (1996), 3247--3250.



\end{thebibliography}
\end{document}